\documentclass[aps,prl,preprint,amssymb,superscriptaddress]{revtex4-2}

\usepackage{url}
\usepackage{hyperref}
\usepackage{graphicx}
\usepackage{amsmath}
\usepackage{siunitx}
\DeclareSIUnit{\photons}{photons}
\DeclareSIUnit{\brilliance}{\photons\per\second\per\mm\squared\per\milli\radian\squared}

\newcommand{\numbril}[1]{\SI{#1}{\brilliance}~0.1\%\,B.W.}

\begin{document}


\title{X-ray Ptychography with a Laboratory Source}


\author{Darren J. Batey}
\affiliation{Diamond Light Source, Harwell Science and Innovation Campus, Fermi Avenue, Didcot, UK}

\author{Frederic Van Assche}
\author{Sander Vanheule}
\author{Matthieu N. Boone}
\affiliation{UGCT-RP, Department of Physics and Astronomy, Ghent University, Belgium}

\author{Andrew J. Parnell}
\affiliation{Soft Matter Analytical Laboratory, Department of Chemistry, University of Sheffield, Sheffield, UK}

\author{Oleksandr O. Mykhaylyk}
\affiliation{Department of Physics and Astronomy, University of Sheffield, Sheffield, UK}

\author{Christoph Rau}
\affiliation{Diamond Light Source, Harwell Science and Innovation Campus, Fermi Avenue, Didcot, UK}

\author{Silvia Cipiccia}
\affiliation{Diamond Light Source, Harwell Science and Innovation Campus, Fermi Avenue, Didcot, UK}
\affiliation{Department of Medical Physics and Biomedical Engineering, University College London, London, UK}

\date{\today}

\begin{abstract}
X-ray ptychography has revolutionised nanoscale phase contrast imaging at
large-scale synchrotron sources in recent years. We present here the first
successful demonstration of the technique in a small-scale laboratory
setting. An experiment was conducted with a liquid metal-jet X-ray source
and a single photon-counting detector with a high spectral resolution. The
experiment used a spot size of 5 µm to produce a ptychographic phase image
of a Siemens star test pattern with a sub-micron spatial resolution. The
result and methodology presented show how high-resolution phase contrast
imaging can now be performed at small-scale laboratory sources worldwide.
\end{abstract}


\maketitle

\section{Introduction}

Ptychography is a coherent scanning-diffraction imaging technique that produces
quantitative images at resolutions beyond the imaging performance of
conventional, lens-based, microscopy systems \cite{ref01}.
Ptychography is now routinely
applied at X-ray synchrotron sources across the world, obtaining highly
sensitive, quantitative, images at the highest spatial resolutions, down to
tens of nanometres \cite{ref02,ref03,ref04,ref05,ref06,ref07}.
Until now, the high level of coherence required for
X-ray ptychography has limited the application of the technique to high
brilliance sources such as synchrotron and, more recently, FEL facilities \cite{ref08}.
It was recently postulated that the new generation of X-ray laboratory sources
may have sufficient brilliance to conduct a ptychographic experiment, given the
correct experimental setup \cite{ref09}.
We present here a demonstration of such an
experiment and the first proof of concept for far field X-ray ptychography
performed using an X-ray laboratory source.

A ptychography scan consists of recording 2D intensity patterns downstream from
a sample that is irradiated by a localised spot of coherent radiation. The 4D
ptychographic dataset is built up by scanning the sample relative to the beam
to a series of overlapping positions. It is possible to record and subsequently
invert the data to retrieve the complex refractive index of the object at
wavelength limited resolutions across an extended field of view \cite{ref10,ref11,ref12,ref13,ref14}. The
success of the inversion step in extracting the phase relies strongly on the
stability of the instrumentation and coherent properties of the beam. The
coherence manifests itself in interference fringes that hold the relative phase
information. The coherent fraction of a beam is related to the lateral (i.e.
spatial) and longitudinal (i.e. temporal) coherence. The former is determined
by the photon energy and the effective source size - how well confined the
source of radiation is laterally in space. The latter is determined by the
source bandwidth - how well confined the source of radiation is in wavelength,
or longitudinally in space. The level of coherence of an instrument can be
described in terms of brilliance. Brilliance is directly proportional to the
spatial and temporal coherence. Typical brilliance of third generation light
sources is of the order of \numbril{e20}

In a recent work, we used a detuned synchrotron source and a hyperspectral
X-ray detector to demonstrate the feasibility of broadband spectroscopic X-ray
ptychography \cite{ref09}. Due to the specific setup, the brilliance of the synchrotron
source was reduced to approximately \numbril{3e11}
State of the art high brilliance X-ray laboratory sources based on a liquid
metal-jet (LMJ) approach this level of brilliance \cite{ref15}. X-ray ptychography
using such a source is therefore feasible, as presented in the following
sections.

\section{Experiment and results}
\subsection{Experimental configuration}

\begin{figure*}
    \includegraphics[width=\linewidth]{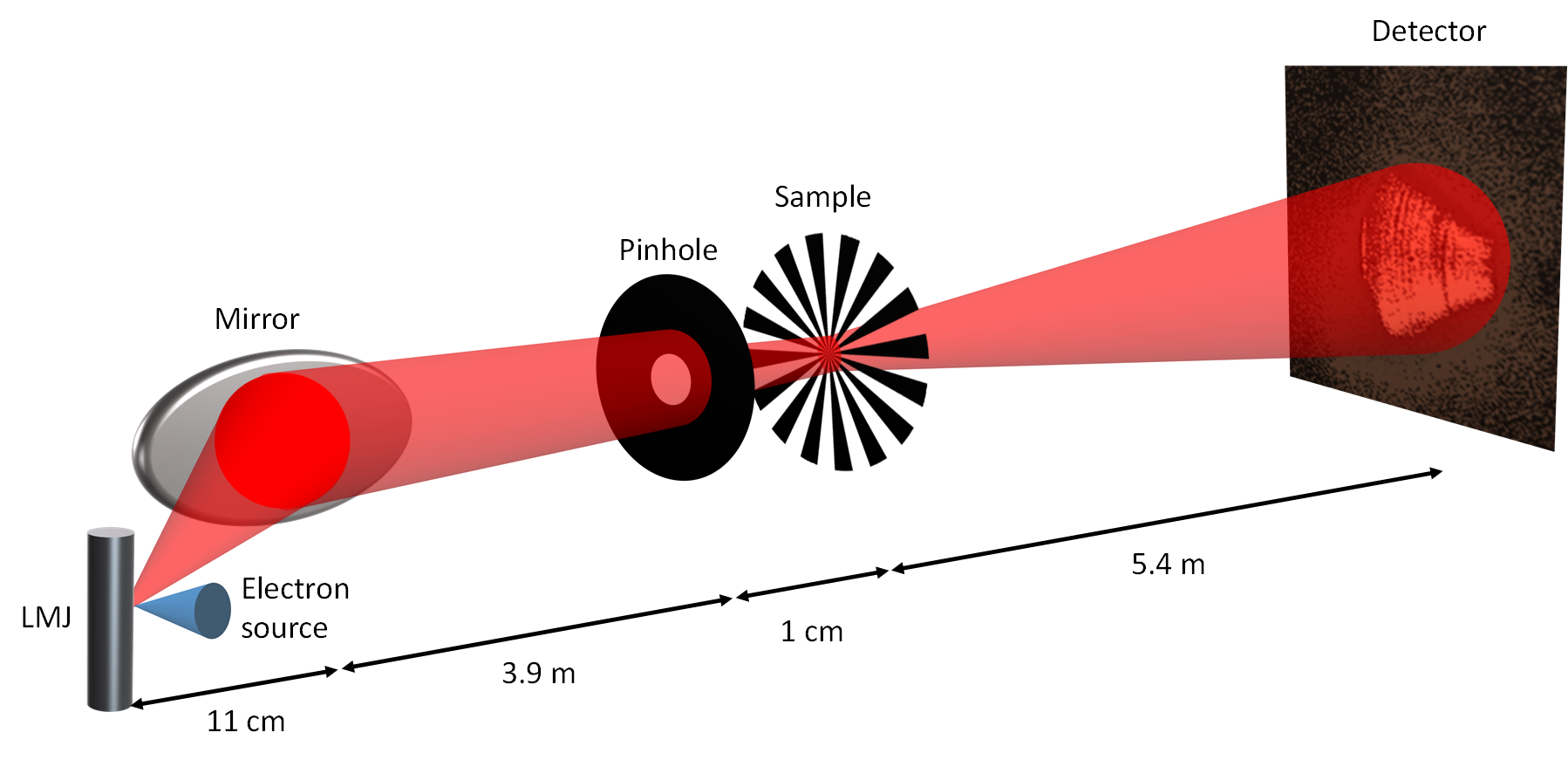}%
\caption{%
    \label{fig:setup}%
    Experimental configuration. Ptychography setup (not to scale) showing the
    experimental layout as implemented at SMALL, Sheffield, UK.}
\end{figure*}

The experiment was designed and conducted to explore the possibility of
ptychographic imaging in a laboratory setting. The data were collected at the
University of Sheffield Soft Matter AnalyticaL Laboratory (SMALL) \cite{ref16,ref17}, with
the portable ptychography end-station from I13-1 of Diamond Light Source and a
hyperspectral detector from Ghent University \cite{ref18,ref19}. The X-ray source is an
Excillum liquid gallium metal jet (LMJ), which has a brilliance of
approximately of \numbril{5e11} \cite{ref15}, one order of
magnitude higher than conventional microfocus sources \cite{ref20,ref21}. The experimental
setup is shown in Figure~\ref{fig:setup}.
The spectrum of the source was characterised using the hyperspectral detector
and it shows the Gallium (Ga) K-alpha and K-beta lines above the Bremsstrahlung
background (see Figure~\ref{fig:spectrum} in method section for additional details).

The experiment consisted of a ptychographic scan of a Siemens star test
pattern. The ptychographic data was reconstructed through the ePIE operator
\cite{ref12} in PtyREX \cite{ref22}, where the set of intensity measurements are inverted into
an image of the object.

\begin{figure*}
    \includegraphics[width=0.8\linewidth]{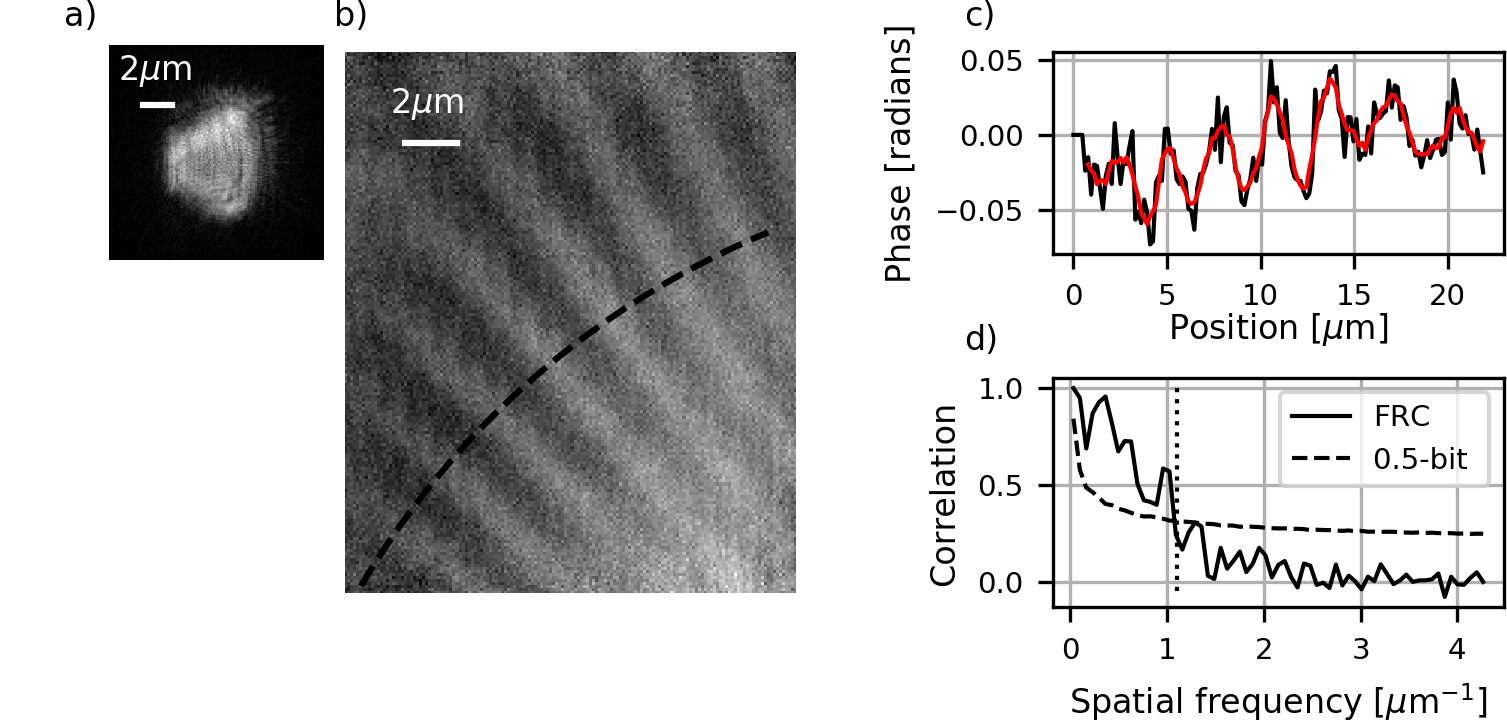}%
\caption{%
    \label{fig:result}%
    The ptychography reconstruction. a) Modulus of the beam profile at the
    sample plane. b) Phase image of the Siemens star test target with a
    reconstructed pixel size of 116~nm. c) Line plot from the dashed black line
    in (b). The raw data is represented by the solid black line and the 10~pixel
    moving average is represented by the solid red line. d) Fourier ring
    correlation of the two split exposure reconstructions showing 1.08~µm$^{-1}$
    spatial frequency (corresponding to 930~nm).}
\end{figure*}

The detector made it possible to post-process the data for different bandwidths
and exposure times. Two datasets, of bandwidth 200~eV (matching the detector
energy resolution at the Ga K-alpha energy) and 1~keV around the K-alpha line, were generated.  The
natural spectral width of the K-alpha lines is a few electronvolt, and the distance
between K-alpha(1) and K-alpha(2) is 17~eV, hence the recorded bandwidth is
determined by the detector energy resolution.
Increasing the bandwidth of the data analysed from 200~eV to 1~keV increases
the contribution of the Bremsstrahlung background. The theoretical resolution
achievable for 200~eV and 1~keV bandwidths are 100~nm and 540~nm respectively \cite{ref23}.
Conversely, the reconstructions of the experimental data showed a decrease in resolution for
the narrower bandwidth (1200~nm for 200~eV and 930~nm for 1~keV), suggesting
that the experiment is photon limited. The best reconstruction,
Figure~\ref{fig:result}(b), was obtained with 1~keV bandwidth.
Both reconstructions included the correction for the
source position and direction. These corrections were essential to compensate
for long-term instabilities of the source during the acquisition (see method
for details). Figure~\ref{fig:result}(a) confirms a beam profile of 5~µm in
extent. A line profile across the reconstructed phase image of the object shows
that the spokes are well resolved (Figure~\ref{fig:result}(c)). The half-bit
resolution of the image is 930~nm (Figure~\ref{fig:result}(d)), a
factor of more than 5 beyond the spot size at the sample.

\subsection{Discussion and summary}
Performing lab-based X-ray ptychography has required advances in lab-sources
\cite{ref20} and detector technologies \cite{ref19}. The high brilliance of the LMJ has
provided the coherent flux required for the ptychography technique. The
hyperspectral detector has been required to characterise spectrally the source
and assess the temporal coherence.

Our analysis of the results suggests that the experiment was limited by the
photon statistics and point-to-point stability of the source. The effect of the
latter was mitigated via the reconstruction algorithm that modelled the source
shift and direction. The use of different or additional optical components for
focussing the X-ray beam to the sample could help to better harness the
coherent flux, increasing the photon statistics and reducing the sensitivity to
long term source instabilities.

We have demonstrated that it is possible to perform X-ray ptychography with a
LMJ source and have shown how to perform ptychography in a laboratory setting,
releasing to the laboratory environment a technique otherwise confined to
synchrotron facilities. The experimental breakthrough achieved with a LMJ is a
first step toward expanding X-ray ptychography to other bright compact light
sources: from inverse Compton scattering \cite{ref24}, to laser-plasma based \cite{ref25} and
compact storage rings \cite{ref26}.

\section{Methods}
\subsection{Experimental configuration}
\paragraph{X-ray source}
The X-ray beam is generated using a JXS-D2-001 liquid metal-jet
laboratory source modified to a higher power performance (Excillum AB, Kista,
Sweden) with Gallium as anode material. The focal spot size of the source can
be varied within a relatively wide range between 5~µm and more than 50~µm, by
tuning the projection of the electron beam on the Gallium jet stream with a set
of electromagnetic lenses. For this experiment, the focal spot size was set to
a nominal value of 5~µm. A three-dimensional single reflection multi-layered
ellipsoidal mirror (FOX3D 11-600 Ga, Xenocs, Grenoble, France) is used to focus
the X-ray beam. The centre of the mirror is located 11~cm downstream of the
X-ray source, coinciding with the first mirror focus. The resulting beam is
slightly converging, with the second mirror focus located approximately 5.2~m
downstream of the mirror. Due to the chromatic behaviour of the mirror
reflectivity, the mirror also acts as a spectral band-pass filter, enhancing
the relative intensity of the 9.25~keV K-alpha emission line of Gallium by
drastically reducing the Bremsstrahlung continuum spectrum.

\paragraph{Scanning system}
The portable ptycho-scope end-station developed at the I13-1
branchline of the Diamond Light Source was used for positioning the sample and
the pinhole (Figure~\ref{fig:setup}).
The ptycho-scope consists of two 3-axis SLC2430 piezo stages (SmarActs GmbH,
Oldenburg, Germany), one for the pinhole and one for the sample. The stages are
controlled with a python data collection software connected to an MCS control
box over an RS232 protocol. The software scans the position point-by-point,
triggers the detector through a USB-BNC connection and uses the detector ready
status for synchronising the motion with the detector readout and beam status.

The instrument was set up with a 5~µm diameter, 50~µm thick, tungsten pinhole
placed 4~m from the source. During the experiment, a flux through the pinhole
of $\sim\SI{5e3}{\photons\per\second}$ was measured.
The sample was placed 1~cm downstream of the pinhole and scanned in the plane
perpendicular to the optical axis of the beam on a square grid of $20\times20$
steps with step size of 1~µm, following a snake-like trajectory. 

\paragraph{X-ray detector}
The pnCCD based Color X-ray Camera (SLcam) \cite{ref27} was used to
measure the diffraction patterns. The detector has a physical pixel pitch of
48~µm, and an active area of $264\times264$ pixels. The system was operated at a
readout speed of 400~fps. The in-house developed software SpeXiDAQ \cite{ref28} was used for
camera control and readout as well as raw data processing. The energy
resolution of the SLcam is approximately 144~eV FWHM at the Mn K-alpha peak and
the centre of mass accuracy is better than 10~eV \cite{ref29}. The detector was placed
downstream of the sample at 9.4~m from the source. Vacuum pipes were placed
between the sample and the detector as well as between the mirror and the
pinhole to reduce the air absorption and scattering. The detector exposure at
each point was 140~s, with a single scan taking 16 hours in total. 

\paragraph{Sample}
The Siemens star is a 500~nm thick gold structure deposited on a silicon
nitride membrane with an outer spoke separation of 4~µm and an inner spoke
separation of 50~nm. An area of 400~µm$^2$ was scanned during the experiment.

\subsection{Data processing}
\begin{figure}
    \includegraphics[width=\linewidth]{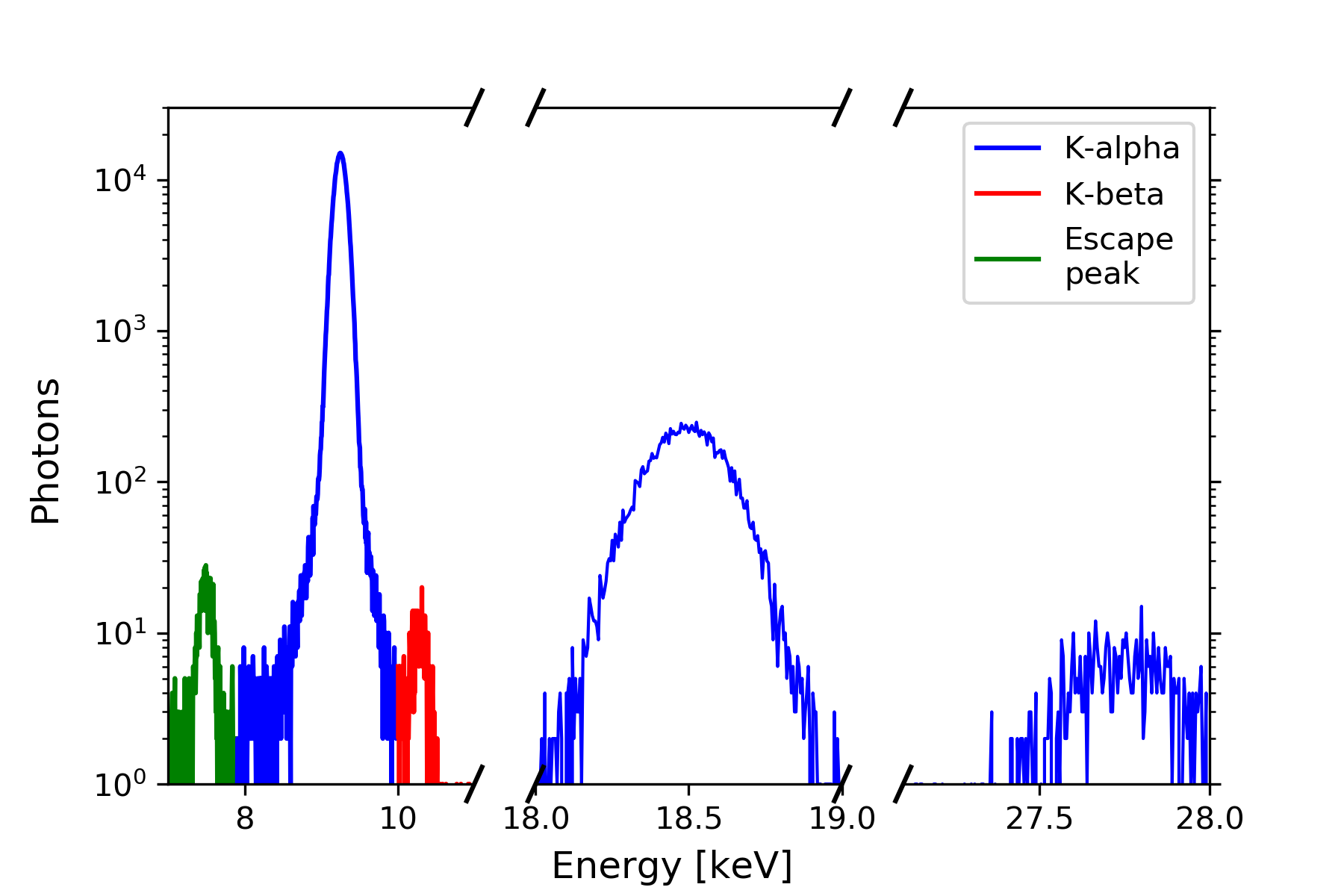}%
\caption{%
    \label{fig:spectrum}%
    The X-ray source spectrum as recorded with the SLcam. The Ga K-alpha peak
    (9.25~keV) along with the double pile-up (18.5~keV) and triple pile-up
    peaks (27.75~keV) are shown in blue. The K-alpha escape peak in Si
    (7.51~keV) is shown in green and the K-beta (10.26~keV) is shown in red.}
\end{figure}

\paragraph{Detector frame processing}
The SLcam captures raw frames containing only a few photon events per frame.
The raw frames are subsequently pre-processed into diffractograms for the
ptychography reconstruction, using a cluster-finding algorithm and subsequent
rebinning of the retrieved events into a 3D datacube (two spatial dimensions
and one spectral dimension). Due to this processing method, charge sharing
effects do not deteriorate the spectral response and sub-pixel accuracy can be
achieved \cite{ref30}. Using SpeXiDAQ \cite{ref28}, the raw frames are stored and afterward
processed and split by time or energy into different datasets. The time-based
splitting was used for assessing the spatial resolution (see post-processing
section below), the energy-based splitting for investigating the spectral
properties. The spectrum has been generated by integrating the photon counts in
each of the 5~eV energy bin datasets. The spectrum recorded is shown in
Figure~\ref{fig:spectrum}: the escape peak of the K-alpha line in the Silicon
bulk and
the double and triple photon pile-up of the K-alpha are visible, beside the
main Ga K-alpha and K-beta peaks. The source spectrum has been retrieved by
adding the counts of the escape peak, those of the double pile-up ($\times 2$) and of
the triple pile-up ($\times 3$) to the K-alpha peak. The energy bandwidth was
investigated and matched to the resolution achievable from the experimental
conditions (flux and geometry).  The data shown in Figure~\ref{fig:result} was
produced using a single output bin ranging from \SIrange{8.75}{9.75}{\keV}.

\subsection{Image reconstruction}
The image reconstruction process takes a model of the experiment, including
knowledge of the illumination conditions and scanning coordinates along with
the recorded intensity measurements, and applies physical constraints in order
to solve for the unknown sample. Here, the illumination was initially modelled
as a convergent beam of 1~mrad full angle and a defocus of 10~mm. The
convergence angle is calculated from the beam on the detector, and the defocus
was chosen to produce a 5~µm spot creating a balance between the true focal
distance of the mirror and the beam width imposed by the pinhole. The scanning
coordinates are taken from the requested values of the SmarAct motors.

\begin{figure*}
    \includegraphics[width=\linewidth]{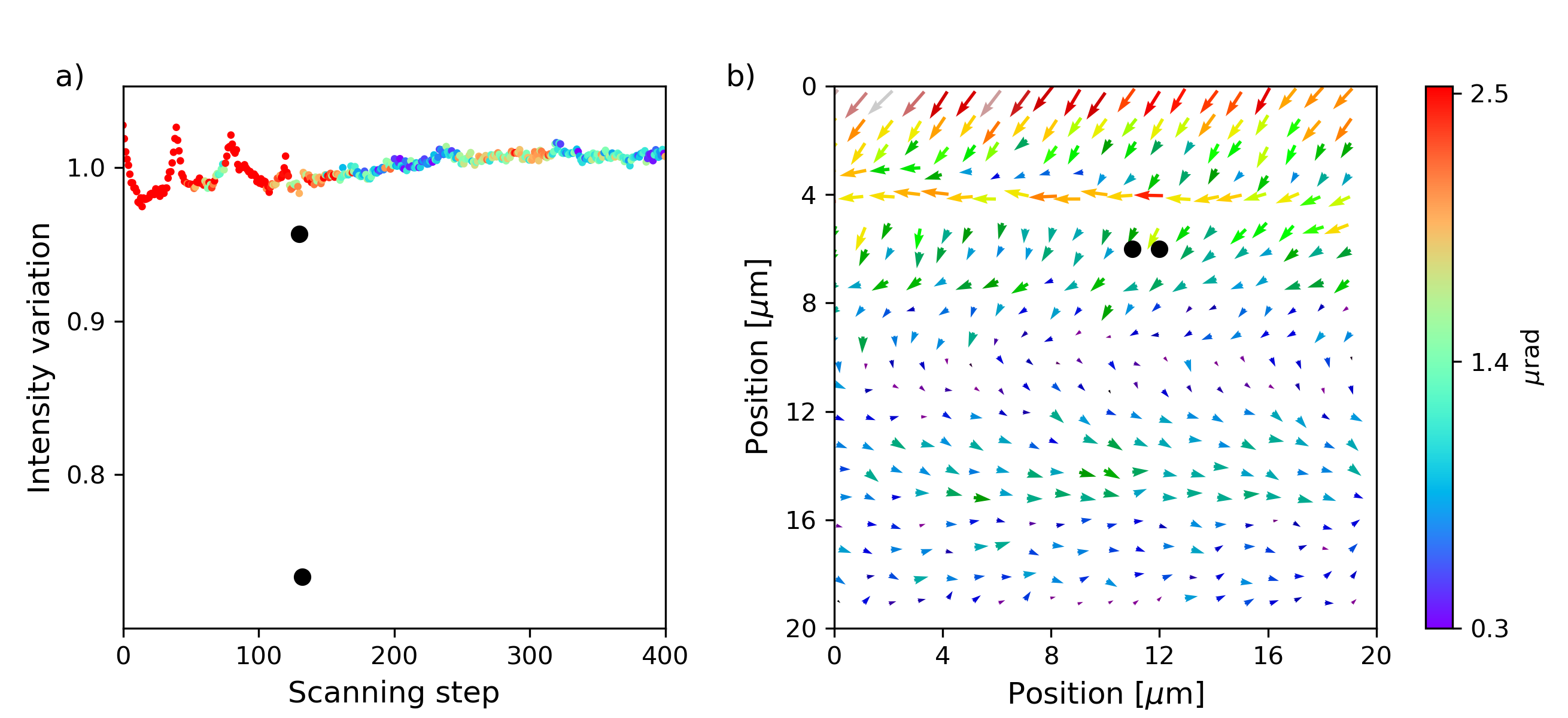}%
\caption{%
    \label{fig:scan}%
    Source and illumination stability. a) Fluctuation of total intensity
    measured by integrating the total flux on the detector during the scan. The
    value is normalised by the mean value. b) Beam position and tilt correction
    as recovered during the PtyREX reconstruction. The origin of the arrow
    represents the re-calculated position, the direction of the arrow
    represents the direction of the tilt correction, and the length of the
    arrow is proportional to the modulus of the angular tilt. The colormap of
    both (a) and (b) represents the modulus of the angular tilt at each scan
    position, highlighting the correlation between intensity fluctuations and
    angular tilt corrections. The two points removed during the reconstruction
    are marked as black dots both in (a) and (b).}
\end{figure*}

The ptychographic data were processed with 500 iterations of the ePIE operator
\cite{ref12} available in PtyREX \cite{ref22}. The reconstruction algorithm is capable of
dealing with source instability, experimental errors and signal degradation due
to noise and decoherence. The beam intensity was monitored during the
acquisition by integrating the flux received on the detector. The intensity
variations, shown in Figure~\ref{fig:scan}(a), are a manifestation of the source
instabilities. The source appears to fluctuate across the first 100 positions,
with a significant sudden drop in intensity at position 131 of the scan. Scan
positions 131 and 132 were removed from the data prior to the reconstruction
(see Figure~\ref{fig:scan}). The source fluctuations translate into point-to-point
instabilities at the sample plane and correspond to either a translation, a
tilt, or a combination of the two in the beam profile. PtyREX employs a scan
correction built on the annealing method of Maiden et al. \cite{ref31}, but is extended
to also accommodate angular variations in the incident beam within the same
update step. The position and tilt correction applied during the reconstruction
are shown Figure~\ref{fig:scan}(b).

The impact of the source properties, detector readout, and beam-sample
positions on the reconstruction quality, was investigated. In order to
understand the effects of each element and to extract the maximum image
quality, a multidimensional parameter sweep was performed on the HPC cluster of
Diamond Light Source. The parameters included were the number of source states
\cite{ref32,ref33}, number of scan correction trials \cite{ref31}, detector threshold levels, and
the bandwidth of the diffraction data. Each parameter permutation was executed
on the split and complete exposure data, allowing for a quantitative comparison
of the resolution.

\paragraph{Post-processing}
To quantify the attained resolution, the acquired dataset was divided (in time)
into two half datasets to perform a Fourier Ring Correlation (FRC) analysis.
The correlation between the two half datasets was compared to a half-bit
information threshold, using an implementation based on van Heel et al. \cite{ref34}.
The splitting was done by alternately assigning a camera time-frame series to
the odd or even dataset. Since the frame interval is very short (2.5~ms)
compared to the expected timescale of source fluctuations, these half datasets
can be considered to be statistically independent measurements of the same
source-object-camera system, including its fluctuations. The obtained FRC curve
is shown in Figure~\ref{fig:result}(d), as well as the half-bit
threshold curve used to determine the attained resolution.
The crossover point of the two curves lies
at 1.08~µm$^{-1}$, corresponding to a resolution of 930~nm. The correlation in
Fourier-space was determined over 65 rings. For completeness, a line profile is
provided, taken along an arc centred at the middle of the Siemens star. 


\begin{acknowledgments}
\section{Acknowledgements}
O.O.M thanks EPSRC for the capital equipment grant (EP/M028437/1) to purchase the laboratory-based Xenocs Xeuss 2.0/Excillum SAXS beamline used for the data collection.

The Research Foundation - Flanders (FWO) is acknowledged for the financial support to this work (Grant number G0A0417N).

The authors acknowledge Dr. Christian David for the design and production of the Siemens star test pattern.

\section{Contribution}
The experiment was conceived by D.J.B. and S.C. and was conducted by D.J.B.,
S.C., F.V.A., S.V., O.O.M and A.J.P.. The portable ptycho-scope was
developed by S.C., D.J.B and C.R.. The raw data were processed by F.V.A.,
S.V and M.N.B.. Reconstruction and analysis was performed by D.J.B., S.V.
and S.C.. The manuscript was written by D.J.B., S.C., M.N.B., S.V. and
F.V.A.. All the authors reviewed the manuscript.
\end{acknowledgments}

\bibliography{bibliography}

\end{document}